\documentclass[11pt]{article}
\usepackage{longtable,supertabular}
\usepackage{amsmath}
\usepackage{amssymb}
\usepackage{latexsym}
\usepackage{cite}

\title{\bf New Binary Self-Dual Cyclic Codes with Square-Root-Like Minimum Distances}
\author{Hao Chen
  \thanks{Hao Chen is with the College of Information Science and Technology/Cyber Security, Jinan University, Guangzhou, Guangdong Province, 510632, China, haochen@jnu.edu.cn. The research of Hao Chen was supported by NSFC Grant 62032009.}}

\begin{document}

\maketitle
\begin{abstract}
The construction of self-dual codes over small fields such that their minimum distances are as large as possible is a long-standing challenging problem in the coding theory. In 2009, a family of binary self-dual cyclic codes with lengths $n_i$ and minimum distances $d_i \geq \frac{1}{2} \sqrt{n_i}$, $n_i$ goes to the infinity for $i=1,2, \ldots$, was constructed. In this paper, we construct a family of (repeated-root) binary self-dual cyclic codes with lengths $n$ and minimum distances at least $\sqrt{n}-2$. New families of lengths $n=q^m-1$, $m=3, 5, \ldots$, self-dual codes over ${\bf F}_q$, $q \equiv 1$ $mod$ $4$,  with their minimum distances larger than or equal to $\sqrt{\frac{q}{2}}\sqrt{n}-q$ are also constructed.\\

{\bf Index terms:} Self-dual code, Dual-containing code, Cyclic code, BCH bound.
\end{abstract}

\section{Introduction and Preliminaries}

The Hamming weight $wt({\bf a})$ of a vector ${\bf a} \in {\bf F}_q^n$ is the number of non-zero coordinate positions of ${\bf a}$. The Hamming distance $d({\bf a}, {\bf b})$ between two vectors ${\bf a}$ and ${\bf b}$ is $d({\bf a}, {\bf b})=wt({\bf a}-{\bf b})$. For a code ${\bf C} \subset {\bf F}_q^n$, its Hamming distance is $$d({\bf C})=\min_{{\bf a} \neq {\bf b}} \{d_H({\bf a}, {\bf b}),  {\bf a} \in {\bf C}, {\bf b} \in {\bf C} \}.$$  The minimum Hamming distance of a linear code is its minimum Hamming weight.  A $[n, k, d]_q$ code over ${\bf F}_q$ is a linear code with the length $n$, the dimension $k$ and the minimum distance $d$. The Singleton bound asserts that $d \leq n-k+1$ for a linear $[n, k, d]_q$ code. A linear code attaining this bound is called  maximal distance separable (MDS). Reed-Solomon codes are well-known MDS codes, see \cite{HP,Lint,MScode}. The Euclidean inner product on ${\bf F}_q^n$ is defined by $$<{\bf x}, {\bf y}>=\Sigma_{i=1}^n x_iy_i,$$ where ${\bf x}=(x_1, \ldots, x_n)$ and ${\bf y}=(y_1, \ldots, y_n)$. The Euclidean dual of a linear code ${\bf C}\subset {\bf F}_q^n$ is $${\bf C}^{\perp}=\{{\bf c} \in {\bf F}_q^n: <{\bf c}, {\bf y}>=0, \forall {\bf y} \in {\bf C}\}.$$ A linear code ${\bf C} \subset {\bf F}_q^n$ is self-orthogonal if ${\bf C} \subset {\bf C}^{\perp}$, is self dual if ${\bf C}={\bf C}^{\perp}$, is dual-containing if ${\bf C}^{\perp} \subset {\bf C}$, and is linear complementary dual (LCD) if ${\bf C}^{\perp} \cap {\bf C}={\bf 0}$. The dual of a dual-containing code is a self-orthogonal code.\\

If $(c_0, c_1, \ldots, c_{n-1}) \in {\bf C}$, then $(c_{n-1}, c_0, \ldots, c_{n-2}) \in {\bf C}$, this code ${\bf C} \subset {\bf F}_q^n$ is called cyclic. A codeword ${\bf c}$ in a cyclic code is identified with a polynomial ${\bf c}(x)=c_0+c_1x+\cdots+c_{n-1}x^{n-1}\in {\bf F}_q[x]/(x^n-1)$. Every cyclic code is a principal ideal in the ring ${\bf F}_q[x]/(x^n-1)$ and then generated by a factor of $x^n-1$. The code with the generator polynomial ${\bf g}(x)=g_0+g_1x+\cdots+g_{n-k}x^{n-k} \in {\bf F}_q[x]$ is denoted by  ${\bf C}_{{\bf g}}$. The dimension of the cyclic code ${\bf C}_{{\bf g}}$ generated by ${\bf g}(x)$ is $n-\deg({\bf g}(x))$.\\

The dual code of a cyclic code ${\bf C}_{{\bf g}}$ is a cyclic code with the generator polynomial ${\bf g}^{\perp}=\frac{x^k{\bf h}(x^{-1})}{{\bf h}(0)}$, where ${\bf h}(x)=\frac{x^n-1}{{\bf g}(x)}$. Therefore the root of ${\bf g}^{\perp}$ is of the form $\frac{1}{\beta}$ if $\beta$ is not a root of ${\bf g}(x)$, where $\beta$ is a $n$-th root of $1$ in some extension field of ${\bf F}_q$, see \cite{HP} Chapter 4.\\

BCH codes were introduced in 1959-1960, see \cite{BC1,BC2,Hoc}, for giving a lower bound on minimum distances of cyclic codes. Duadic codes were introduced in 1981 by Leon, Pless and Sloane, see \cite{Leon}, as extensions of quadratic residue codes. It is well-known that duadic cyclic codes have the square root lower bound on their minimum distances, see \cite[Chapter 6]{HP}. In three recent papers \cite{TangDing,SunDW,SunDing}, new binary cyclic codes, new negacyclic ternary codes and new constacyclic codes with the square-root-like  lower bound on their minimum distances and dual minimum distances were constructed.\\

For self-dual codes over small fields with small lengths, we refer to \cite{CPS,CS,DGH,Rains} and \cite{HP} Chapter 9.  It is always interesting in the coding theory to construct self-dual codes with large minimum distances. It was proved that there is a family of binary self-dual codes meeting the Gilbert-Varshamov bound, and there is a family of self-dual codes over ${\bf F}_q$ exceeding the Gilbert-Varshamov bound for $q \geq 64$ and $q \neq 125$, see \cite{Rains,Bassa}. In the finite length regime, there have been many papers constructing self dual generalized Reed-Solomon (GRS, then MDS) codes, we refer to \cite{GG08,Gulliver,JX17,Fang,Kim,ZhangFeng} and references therein. The minimum weight $d$ of a length $n$ binary self-dual code satisfies $$d\leq 4 \lfloor \frac{n}{24}\rfloor+6,$$ if $n \equiv 22$ $mod$ $24$, and $$d \leq 4 \lfloor \frac{n}{24}\rfloor+4,$$ otherwise. A binary self-dual code attains this upper bound is called an extremal self-dual code, see e.g. \cite[Chapter 19]{MScode}. The minimum weight $d$ of a length $n$ ternary self-dual code satisfies $$d\leq 3 \lfloor \frac{n}{12}\rfloor+3.$$ A ternary self-dual code attains this bound is called an extremal self-dual ternary code. There has been a long history in the coding theory to construct self-dual codes over small fields with small lengths and large minimum distances, we refer to \cite{DGH,Gulliver1,Harada,Gaborit,Gaborit1,Gaborit3,Dijk,Gulliver2,Bet,Shi}. Many authors have constructed self-dual codes over ${\bf F}_q$ with various lengths for small prime powers $q$. On the opposite direction, it was proved in \cite{Carlet} that each linear code over ${\bf F}_q$, $q >3$, is equivalent to an LCD code. The hull-increasing variation problem of equivalent linear codes was proposed and studied in \cite{Chen}.\\

It is well-known that extended (binary) quadratic residue codes (then an extended code of a cyclic code) with length $p+1$ and the minimum distance $\sqrt{p}$ are self-dual, where $p \equiv \pm 1$ $mod$ $8$.  Some extended duadic cyclic codes are self-dual codes with square-root minimum distances, see \cite[Chapter 6]{HP}, \cite{Leon}. It was proved that self-dual codes over ${\bf F}_q$ only exist when $q$ is even, see \cite{Kai,Jia}.\\

Self-dual cyclic codes can be thought as self-dual codes with large automorphism groups. Thus it is interesting to construct self-dual cyclic codes or negacyclic codes over small fields with large minimum distances. In 2009, there was an important progress that a family of binary self-dual cyclic codes with the length $n=2(2^{2a+1}-1)$ and the minimum distance $\delta \geq \frac{1}{2} \sqrt{n+2}$ were constricted, for $a=1,2 \ldots$, by B. Heijne and J. Top in  \cite{HTop}. To the best of our knowledge, there is no other construction of binary self-dual cyclic codes with a large lower bound on minimum distances, except \cite {HTop}. It is natural to ask if new families of binary self-dual cyclic codes with larger minimum distances can be constructed.\\

In this paper, we give the $[{\bf u}|\lambda{\bf u}+{\bf v}]$ construction of self-dual codes from dual-containing codes. Then a family of self-dual codes over ${\bf F}_q$, $q=2$ or $q\equiv 1$ $mod$ $4$, with the length $n=\frac{2(q^m-1)}{\mu}$, $\mu$ is a divisor of $q^m-1$,  and the minimum distance $\sqrt{\frac{q}{2\mu}}\sqrt{n}-q$ is constructed. From the classical result due to van Lint in \cite{Lint1}, these constructed self-dual codes over ${\bf F}_2$ are repeated-root binary cyclic codes. Then our construction improves binary self-dual cyclic codes in \cite{HTop} significantly. In the case that $q$ is an odd prime power satisfying $q \equiv 1$ $mod$, $4$, our self-dual codes with lengths $n=\frac{2(q^m-1)}{\mu}$ and minimum distances at least  $\sqrt{\frac{q}{2\mu}}\sqrt{n}-\frac{q}{\mu}$ have the following property, the shift to the right of two positions of any codeword is still a codeword. Therefore self-dual codes constructed in this paper have large automorphism groups.\\

\section{The $[{\bf u}|\lambda{\bf u}+{\bf v}]$ construction of self-dual codes}

We first give the main construction of self-dual codes over ${\bf F}_q$ from dual-containing codes.\\

{\bf Theorem 2.1.} {\em Let ${\bf C} \subset {\bf F}_q^n$ be a dual-containing code. If there is a nonzero element $\lambda \in {\bf F}_q$ satisfying $\lambda^2 =-1$. Then the code $${\bf C}_1=\{{\bf u}|\lambda{\bf u}+{\bf v}: {\bf u} \in {\bf C}, {\bf v} \in {\bf C}^{\perp}\} \subset {\bf F}_q^{2n}$$ is a self-dual code with the minimum distance at least $\min\{d({\bf C}^{\perp}), 2d({\bf C})\}$.}\\

{\bf Proof.} For two codewords ${\bf c}_1=[{\bf u}_1|\lambda{\bf u}_1+{\bf v}_1]$ and ${\bf c}_2=[{\bf u}_2|\lambda{\bf u}_2+{\bf v}_2]$ in this code ${\bf C}_1$, their inner product is $<{\bf c}_1,{\bf c}_2>=(\lambda^2+1)<{\bf u}_1, {\bf u}_2>+\lambda(<{\bf u}_1, {\bf v}_2>+<{\bf u}_2,{\bf v}_1>)+<{\bf v}_1, {\bf v}_2>=0$, since $\lambda^2+1=0$ and ${\bf C}^{\perp}$ is self-orthogonal. It is clear that the dimension of ${\bf C}_1$ is $n$. The conclusion follows immediately.\\

In the case $q=2$, $\lambda=1$. Suppose that the length of ${\bf C}$ is odd and the generator polynomial of ${\bf C}$ is $g_1(x) \in {\bf F}_2[x]$. Since ${\bf C}^{\perp} \subset {\bf C}$, then the generator polynomial of ${\bf C}^{\perp}$ is $g_1(x)g_2(x) \in {\bf F}_2[x]$. From the classical result in \cite{Lint1}, the above construction is a repeated-root cyclic code with the generator polynomial $g_1^2(x) g_2(x)$ after a coordinate permutation. Therefore if ${\bf C}$ in Theorem 2.2 is a binary cyclic code with the odd length, then the code ${\bf C}_1$ is a repeated-root binary self-dual cyclic code. In the case $q$ is odd, when ${\bf C}$ is a cyclic code, it is clear that the group ${\bf Z}_n \times {\bf Z}_n$ is in the automorphism group of ${\bf C}_1$.\\

{\bf Corollary 2.1.} {\em Let $q$ be an even prime power and $n$ be a positive integer satisfying $n \leq q$. Then a self-dual $[2n, n, d\geq \frac{2n}{3}]_q$ code can be constructed from the Reed-Solomon code.}\\

{\bf Proof.} Since $q$ is an even prime power, each element in ${\bf F}_q$ is a square. Then each Reed-Solomon $[n, k, n-k+1]_q$ code is equivalent to a self-orthogonal code when $k \leq \frac{n}{2}$, see \cite{GG08}. Then the conclusion follows from Theorem 2.1 directly by setting $\lambda=1$.\\

\section{Dual-containing BCH codes with $\sqrt{\frac{q}{2\mu}}\sqrt{n}-\frac{q}{\mu}$ minimum distances}

Set ${\bf Z}_n={\bf Z}/n{\bf Z}=\{0, 1, \ldots, n-1\}$. A subset $C_i$ of ${\bf Z}_n$ is called a cyclotomic coset if $$C_i=\{i, iq, \ldots, iq^{l-1}\},$$ where $i \in {\bf Z}_n$ is fixed and $l$ is the smallest positive integer such that $iq^l \equiv i$ $mod$ $n$. It is clear that cyclotomic cosets corresponds to irreducible factors of $x^n-1$ in ${\bf F}_q[x]$. Therefore a generator polynomial of a cyclic code is the product of several irreducible factors of $x^n-1$. The defining set of a cyclic code generated by ${\bf g}$  is the the following set $${\bf T}_{{\bf g}}=\{i: {\bf g}(\beta^i)=0\}.$$ Then the defining set of a cyclic code is the disjoint union of several cyclotomic cosets. Set $${\bf T}^{-1}=\{-i: i \in {\bf T}\}.$$ Then the defining set of the dual code is $({\bf T}^c)^{-1}$, where ${\bf T}^c={\bf Z}_n-{\bf T}$ is the complementary set. It is well-known that a cyclic code with the defining set ${\bf T}$ is dual-containing if and only if ${\bf T} \cap {\bf T}^{-1}=\emptyset$, see \cite{LDL2}.\\

The following lemma is useful in this paper.\\

{\bf Lemma 3.1 (see \cite{LDL2}).} {\em Let $q>1$ be a positive integer and $a,b$ be two positive integers, then $\gcd(q^a-1, q^b-1)=q^{\gcd(a,b)}-1$. $\gcd(q^a+1, q^b-1)=1$ if $q$ is even and $\frac{b}{\gcd(a,b)}$ is odd, $\gcd(q^a+1, q^b-1)=2$ if $q$ is odd and $\frac{b}{\gcd(a,b)}$ is odd, $\gcd(q^a+1, q^b-1)=q^{\gcd(a,b)}+1$ if $\frac{b}{\gcd(a,b)}$ is even.}\\

The following result follows from the BCH bound and the characteristic of dual-containing cyclic codes.\\

{\bf Theorem 3.1.} {\em Suppose that there is no integer $a$ in ${\bf Z}_n$ such that both $a$ and $-a$ in the same cyclotomic coset. Moreover there is no integer $a \in {\bf Z}_n$ satisfying that both $a$ and $-a$ is in the set ${\bf T}=C_1 \cup C_2 \cup \cdots \cup C_{\delta-1}$. Then we can construct a length $n$ dual-containing cyclic code with the defining set ${\bf T}$, such that its minimum distance is at least $\delta$.}\\

{\bf Proof.} First of all from the condition that there is no integer $a \in {\bf Z}_n$ such that both $a$ and $-a$ are in the same cyclotomic coset, all cyclotomic cosets can be paired. Since there is no integer $a \in {\bf Z}_n$, such that $-a$ and $a$ are in the defining set ${\bf T}=C_1 \cup C_2 \cup \cdots \cup C_{\delta-1}$, the cyclic code with this defining set is a dual-containing code. \\

Let $\mu $ be a divisor of $q^m-1$, $m$ odd, and the length be $n=\frac{q^m-1}{\mu}$. If there are $a$ and $-a$ in the same cyclotomic coset, then $a(q^i+1) \equiv 0$ $mod$ $n$, where $i \leq m$. It is easy to verify that this is not possible from Lemma 3.1. We have the following result.\\

{\bf Theorem 3.2.} {\em Let $n$ as above be the length. Then we can construct the BCH dual-containing code ${\bf C}$ with the defining set ${\bf T}={\bf T}=C_1 \cup C_2 \cup \cdots \cup C_{\delta-1}$, where $\delta = \frac{q^{\frac{m+1}{2}}-q}{\mu}$. Then we have $$d({\bf C}) \geq \frac{q^{\frac{m+1}{2}}-q}{\mu}+1.$$}\\

{\bf Proof.} We only need to prove that there is no two positive integer $1 \leq u, v \leq  \frac{q^{\frac{m+1}{2}}-q}{\mu}$, satisfying $uq^t+v \equiv 0$ $mod$ $n$, where $t \leq m$. Without the loss of the generality, we can assume that $t \leq \frac{m-1}{2}$. Otherwise $t \geq \frac{m+1}{2}$, $uq^m+vq^{m-t}\equiv u+vq^{m-t} \equiv 0$ $mod$ $n$. Then $\mu(uq^t+v) \geq q^m-1$, the conclusion follows directly.\\

\section{Binary self-dual cyclic codes with large minimum distances}

From Theorem 2.1 and Theorem 3.2 in the case $q=2$ and $\mu=1$, we have the following construction of binary self-dual cyclic codes with square-root-like minimum distances.\\

{\bf Theorem 4.1.} {\em Let $n=2(2^m-1)$, where $m=3, 5, 7, \ldots$. We can construct a family of binary self-dual (repeated-root) cyclic codes with the length $n$ and the minimum distances at least $\sqrt{n}-2$.}\\

{\bf Proof.} From Theorem 3.2 we have a dual-containing BCH code ${\bf C} \subset {\bf F}_2^{2^m-1}$ with the minimum distance at least $2^{\frac{m+1}{2}}-2$. Then from the construction in Theorem 2.1, the code ${\bf C}_1 \subset {\bf F}_2^{2(2^m-1)}$ is self-dual. The minimum distance of ${\bf C}_1$ is at least $\sqrt{n}-2$. From the classical result in \cite{Lint1} as explained in Section 2, ${\bf C}_1$ is a binary repeated-root cyclic code. The conclusion follows directly.\\

The following result follows from Theorem 2.1 and Theorem 3.2 directly.\\

{\bf Corollary 4.1.} {\em Let $n=\frac{2(2^m-1)}{\mu}$, where $m=3, 5, 7, \ldots$ and $\mu$ be a divisor of $2^m-1$. We can construct a family of binary self-dual (repeated-root) cyclic codes with the length $n$ and the minimum distance at least $\sqrt{\frac{n}{\mu}}-\frac{2}{\mu}$.}\\

{\bf Example 4.1.} We can construct a binary BCH dual-containing $[15, 11, 3]_2$ code. The dual code is a binary cyclic $[15, 4, 7]_2$ code. Then from Theorem 2.1, a binary self-dual cyclic $[30, 15, 6]_2$ code is constructed. The upper bound of the minimum distance of binary self-dual code of the length $30$ is $6$, see \cite{Gaborit2}. Thus when the length is $30$, the minimum distance of an extremal binary self-dual code is the same as the minimum distance of a binary self-dual cyclic code.\\

{\bf Example 4.2.} We can construct a binary BCH dual-containing $[63, 42, 7]_2$ code. The dual code is a binary cyclic $[63, 21, 16]_2$ code. Then from Theorem 2.1, a binary self-dual cyclic $[126, 63, 14]_2$ code is constructed. The upper bound of the minimum distance of binary self-dual code of the length $126$ is $24$, see \cite{Gaborit2}. The constructed length $126$ binary self-dual code with the known largest minimum weight is such a code with the minimum weight $18$. Then it seems that the method in this paper gives a quite good binary self-dual code of the length $126$, with the cyclic structure.\\

These binary self-dual cyclic codes with small lengths from Theorem 2.1, can be compared with binary self-dual codes without the cyclic structure in \cite{DGH,Gaborit2} and references therein.\\

\section{Self-dual codes over ${\bf F}_q$ with $\sqrt{\frac{q}{2}}\sqrt{n}-q$ minimum distances}

Let $q$ be an odd prime power satisfying $q \equiv 1$ $mod$ $4$. Then in the field ${\bf F}_q$, $-1$ is an square. We have the following result from Theorem 2.1 and Theorem 3.2.\\

{\bf Theorem 5.1.} {\em Let $q$ be an odd prime power satisfying $q \equiv 1$ $mod$ $4$. Let $n=\frac{2(q^m-1)}{\mu}$, where $m$ is odd and $\mu$ is a divisor of $q^m-1$. Then a family of self-dual codes over ${\bf F}_q$ of the length $n$ and the minimum distance at least $\sqrt{\frac{q}{2\mu}}\sqrt{n}-\frac{q}{\mu}$ is constructed.}\\

In the case $\mu=1$ we have the following result.\\

{\bf Corollary 5.1.} {\em Let $q$ be an odd prime power satisfying $q \equiv 1$ $mod$ $4$. Let $n=2(q^m-1)$, where $m$ is odd and $\mu$ is a divisor of $q^m-1$. Then a family of self-dual codes over ${\bf F}_q$ of the length $n$ and the minimum distance at least $\sqrt{\frac{q}{2}}\sqrt{n}-q$ is constructed.}\\

The automorphism group of the code constructed in Theorem 5.1 and Corollary 5.1 contains the subgroup ${\bf Z}_{n/2} \times {\bf Z}_{n/2}$.\\

\section{Conclusions}

The construction of binary self-dual codes or self-dual codes over small fields such that their minimum distances are as large as possible is an active topic in the coding theory. In this paper we constructed a family of binary self-dual cyclic codes with the square-root-like minimum distances, which improved the binary self-dual cyclic codes in \cite{HTop} significantly.  New families of self-dual codes over ${\bf F}_q$, $q \equiv 1$ $mod$ $4$,  with lengths $2n$ and  square-root-like minimum distances, and the automorphism group containing the group ${\bf Z}_n \times {\bf Z}_n$, were also constructed. New families of self-dual negacyclic codes with square-root-like minimum distances are constructed in \cite{XieChen}.\\


\begin{thebibliography}{10}

\bibitem{Gaborit3} C. Aguilar Melchor and P. Gaborit, On the clssification of extremal $[36, 18, 8]$  binary self-dual codes, IEEE Trans. Inf. Theory, vol. 54, no. 10, pp. 4743-4750, 2008.

\bibitem{Bassa} A. Bassa and H. Stichtenoth, Self-dual codes better than the Gilbert-Varshamov bound, Des., Codes and Cryptogr., vol. 87, pp. 173-182, 2019.


\bibitem{Bet} K. Betsumiya, S. Georgiou, T. A. Gulliver, M. Harada and C. Koukouvinos, Self-dual codes over some prime fields, Disc. Math., vol. 262, pp. 37-58, 2003.



\bibitem{BC1} R. C. Bose and D. K. Ray-Chaudhuri, On a class of error-correcting binary group codes, Inform. and Control, vol. 3, pp. 68-79, 1960.

\bibitem{BC2} R. C. Bose and D. K. Ray-Chaudhuri, Further results on error-correcting binary group codes, Inform. and Control, vol. 3, pp. 279-290, 1960.







\bibitem{Chen} H. Chen, On the hull-variation problem of equivalent linear codes, IEEE Trans. Inf. Theory, vol. 69, no. 5, pp. 2911-2922, 2023.


\bibitem{Carlet} C. Carlet, S. Mesnager, C. Tang, Y. Qi, and R. Pellikaan, Linear codes over ${\bf F}_q$ are equivalent to LCD codes for $q>3$, IEEE Trans. Inf. Theory, vol. 64, no. 4, pp. 3010-3017, 2018.


\bibitem{CPS} J. H. Conway, V. Pless and N. J. A. Sloane, Self-dual codes over $GF(3)$ and $GF(4)$ of length not exceeding $16$, IEEE Trans. Inf. Theory, vol. 25, no. 3, pp. 312-322, 1979.

\bibitem{CS} J. H. Conway and N. J. A. Sloane, A new upper bound on the minimal distance of self-dual codes, IEEE Trans. Inf. Theory vol. 36, pp. 1319-1333, 1990.


\bibitem{Dijk} M. van Dijk, S. Egner, M. Greferath and A. Wassermann, On two doubly even self-dual binary codes of length 160 and minimum weight 24, IEEE Trans. Inf. Theory vol. 51, no. 1 pp. 408-410, 2005.



\bibitem{DGH} S. T. Dougherty, T. A. Gullier and M. Harada, Extremal binary self-dual codes, IEEE Trans. Inf. Theory vol. 43, no. 6, pp. 2036-2047, 1997.



\bibitem{Fang} W. Fang, New constructions of MDS Euclidean self-dual codes from GRS codes and extended GRS codes, IEEE Trans. Inf. Theory, vol. 65. no. 9, pp. 5574-5579, 2019.

\bibitem{Gaborit2} P. Gaborit, Tables of self-dual codes, Tables de codes auto-duaux, http://www.unilim.fr/pages-perso/phillie.gaborit.



\bibitem{Gaborit1} P. Gaborit, V. Pless, P. Sole and and O. Atkin, Type II codes over ${\bf F}_4$, Finite Fields Appl., vol. 8, pp. 171-183, 2002.



\bibitem{Gaborit} P. Gaborit and A. Otmani, Experimental constructions of self-dual codes, Finite Fields Appl., vol. 9, pp. 372-394, 2003.


\bibitem{Gildea} J. Gildea, A. Korban and A. M. Roberts, New binary self-dual codes of length $80, 84$ and $96$ from composite matrices, Des., Codes and Cryptogr., vol. 90, pp. 317-342, 2022.

\bibitem{GG08} M. Grassl and T. A. Gulliver, On self-dual MDS codes, Proc. Int. Symp. Inf. Theory, pp. 1954-1957, 2008.


\bibitem{Guenda} K. Guenda and T. A. Gulliver, Self-dual repeated root cyclic and negacyclic codes over finite fields, arXiv: 1207.3387,2012.


\bibitem{Gulliver2} T. A. Gulliver, Optimal double circulant self-daul codes over ${\bf F}_4$, IEEE Trans. Inf. Theory, vol. 46, no. 9, pp. 271-274, 2000.


\bibitem{Gulliver} T. A. Gulliver, J-L. Kim and Y. Lee, New MDS or near MDS codes, IEEE Trans. Inf. Theory, vol. 54, no. 9, pp. 4354-4360, 2008.

\bibitem{Gulliver1} T. A. Gulliver and M. Harada, New binary self-dual codes, IEEE Trans. Inf. Theory, vol. 54, no. 1, pp. 415-417, 2008.

\bibitem{Harada} M. Harada, W. Holzmann, H. Kharaghani and M. Khorvash, Extremal ternary self-dual codes constructed from negacirculant matrices, Graphs and Combinat., vol. 23, pp. 401-417, 2007.


\bibitem{HTop} B. Heijne and J. Top, On the minimum distance of binary self-dual cyclic codes, IEEE Trans. Inf. Theory, vol. 55, no. 11, pp. 4860-4863, 2009.

\bibitem{Hoc} A. Hocquenghem, Codes correcteurs d'erreurs, Chiffres (Paris), vol. 2, pp. 147-156, 1959.



\bibitem{HP} W. C. Huffman and V. Pless, Fundamentals of error-correcting codes, Cambridge University Press, Cambridge, U. K., 2003.

\bibitem{Kai} X. Kai and S. Zhu, On cyclic self-dual codes, Appl. Algebra Engr.Commun. Comput., vol. 19, pp. 509-525, 2008.


\bibitem{KS} A. Krishna and V. Sarwate, Pseudocyclic maximal-distance-separable codes, IEEE Trans. Inf. Theory, vol. 36, no. 4, pp. 880-884, 1990.

\bibitem{Jia} Y. Jia, S. Ling and C. Xing, On self-dual cyclic codes over finite fields, IEEE Trans. Inf. Theory, vol. 57, no. 4, pp. 2243-2251, 2011.


\bibitem{JX17}  L. Jin and C. Xing, New MDS self-dual codes from generalized Reed-Solomon codes,  IEEE Trans. Inf. Theory, vol. 63, no. 3, pp. 1434-1438, 2017.

\bibitem{Leon} J. S. Leon, V. Pless and N. J. A. Sloane, Duadic codes, IEEE Trans. Inf. Theory, vol. 30, pp. 709-714, 1981.


\bibitem{LDL2} C. Li, C. Ding and S. Liu, LCD cyclic codes over finite fields, IEEE Trans. Inf. Theory, vol. 63, no. 7, pp. 4344-4356, 2017.




\bibitem{Lint} J. H. van Lint, Introduction to the coding theory, GTM 86, Third and Expanded Edition, Springer, Berlin, 1999.

\bibitem{Lint1} J. H. van Lint, Repeated-root cyclic codes, IEEE Trans. Inf. Theory, vol. 37, no. 2, pp. 343-345, 1991.

\bibitem{Kim} J.-L. Kim and Y. Lee, Euclidean and Hermitian self-dual MDS codes over large finite fields, J. Combinat. Theory, A, vol. 105, no. 1, pp.79-95, 2004.



\bibitem{MScode} F. J.  MacWilliams and N. J. A. Sloane, The Theory of error-correcting codes, 3rd Edition, North-Holland Mathematical Library, vol. 16. North-Holland, Amsterdam, 1977.





\bibitem{Rains} E. M. Rains and N. J. A. Sloane, Self-dual codes, In "Handbook of Coding Theory", eds, V. Pless and W. C. Huffman, pp. 177-294, Elsevier, Amsterdam, 1998.


\bibitem{Shi} M. Shi, L. Sok, P. Sole and S. Calkavur, Self-dual codes and orthogonal matrices over large finite fields, Finite Fields Appl., vol. 54, pp. 297-314, 2018.

\bibitem{SunDing} Z. Sun and C. Ding, Several families of ternary negacyclic codes and their duals, arXiv:2301.09783vs, 2023.

\bibitem{SunDW} Z. Sun, C. Ding and X. Wang, Two classes of constacyclic does with variable parameters, arXiv:2208.05664v3, 2022.

\bibitem{TangDing} C. Tang and C. Ding, Binary $[n, \frac{n+1}{2}]$ cyclic codes with good minimum distances, IEEE Trans. Inf. Theory, vol. 68, no. 12, pp. 7842-7849, 2022.




\bibitem{XieChen} C. Xie and H. Chen, New families of self-dual negacyclic codes with various lengths and square-root-like minimum distances, preprint, 2023.


\bibitem{ZhangFeng} A. Zhang and K. Feng, A unified approch to construct MDS self-dual codes via Reed-Soloon code, IEEE Trans. Inf. Theory, vol. 66, no. 6, pp. 3650-3656, 2020.






\end{thebibliography}
\end{document}